\documentclass[prd,nofootinbib,showkeys,showpacs,preprint]{revtex4}
\usepackage{amsmath,graphicx}
\usepackage{booktabs}

\begin{document}

\title{The rho meson in a scenario of pure chiral restoration}
\author{T. Hilger, R. Thomas, B. K\"ampfer}
\affiliation{
Forschungszentrum Dresden-Rossendorf, PF 510119, D-01314 Dresden, Germany\\
TU Dresden, Institut f\"ur Theoretische Physik, 01062 Dresden, Germany}
\author{S. Leupold}
\affiliation{Institutionen f\"or fysik och astronomi, Uppsala Universitet, Sweden}

\begin{abstract}
Based on QCD sum rules we explore the consequences of 
a pure chiral restoration scenario for the $\rho$ meson, 
where all chiral symmetry breaking condensates are dropped 
whereas the chirally symmetric condensates remain at 
their vacuum values. This pure chiral restoration scenario
causes the drop of the $\rho$ spectral moment by about
120 MeV. 
The complementarity of mass shift and broadening is discussed.
A simple parametrization of the $\rho$ spectral function leads to 
a width of about 600 MeV if no shift of the peak position is assumed.
\end{abstract}

\maketitle

{\bf 1. Introduction:}
The impact of chiral symmetry restoration on the properties of hadrons is a much
debated issue. In particular light vector mesons have been studied extensively
both on the theoretical and the experimental side; for recent reviews see e.g.\
\cite{Rapp:2009yu,Rapp:1999us,Tserrua,Metag,Leupold:2009kz}. 
In fact, in-medium modifications of hadrons 
made out of light quarks and especially their possible ''mass drops''
are taken often synonymously for chiral restoration. 
The Brown-Rho scaling conjecture \cite{Brown:1991kk} and
Ioffe's formula for the nucleon \cite{Ioffe:1981kw}
suggest such a tight connection.
However, experimentally the main observation of in-medium
changes of light vector mesons via dilepton spectra is a significant broadening 
of the spectral shape \cite{Arnaldi:2006jq,Adamova:2006nu}. 
Such a broadening can be obtained in hadronic
many-body approaches, e.g., 
\cite{Friman:1997tc,Rapp:1997fs,Peters:1997va,Post:2003hu,vanHees:2006ng}, 
which at first sight are not related directly to chiral restoration
in the above spirit. 
Pion dynamics and resonance formation, both fixed to vacuum data, provide the important
input for such many-body calculations. Clearly the pion dynamics is closely linked to
the vacuum phenomenon of chiral symmetry {\em breaking}, but the connection to
chiral {\em restoration} is not so clear. For the physics of resonances the connection
is even more loose. There are recent approaches which explain some hadronic
resonances as dynamically generated from chiral 
dynamics \cite{Kaiser:1995cy,Kolomeitsev:2003kt,Sarkar:2004jh,Lutz:2003fm,%
Roca:2005nm,Wagner:2008gz,Wagner:2007wy}, but again this
primarily points towards an intimate connection between hadron physics and
chiral symmetry breaking and not so much chiral restoration. As suggested, e.g., in
\cite{Rapp:1999us,Leupold:2008ne} the link to chiral restoration might be indirect: The
in-medium broadening could be understood as a step towards deconfinement. In the deconfined
quark-gluon plasma also chiral symmetry is presumed to be restored.
All these considerations suggest
that the link between chiral restoration and in-medium changes of hadrons is not
as clear as one might have hoped.

Additional input could come from approaches which are closer to QCD than standard
hadronic models. One such approach is the QCD sum rule method 
\cite{Shifman:1978bx,Shifman:1978by,Reinders:1984sr,Hatsuda:1991ez,Hatsuda:1992bv}.
A somewhat superficial view on QCD sum rules for vector mesons seems to support
the original picture of an intimate connection between chiral restoration and 
in-medium changes. Here the previously popular chain of arguments goes as follows:
(1) Four-quark condensates
play an important role for the vacuum mass of the light vector 
mesons \cite{Shifman:1978bx,Shifman:1978by} (see, however, \cite{Kwon:2008vq} 
for a different view).
(2) The four-quark condensates factorize into squares of the two-quark 
condensate \cite{Novikov:1983jt}.
(3) The two-quark condensate drops in the medium due to chiral 
restoration \cite{Gerber:1988tt,Drukarev:1991fs}.
(4) Thus the four-quark condensates drop in the medium accordingly.
(5) Therefore the masses of light vector mesons change (drop) in the medium due to 
chiral restoration.

In this line of reasoning only the points 1 and 3 are undoubted. Even if one follows
the arguments of points 1 to 4 it has been shown that besides a dropping mass also
a broadened hadronic spectral distribution is compatible with the QCD sum rules
\cite{Klingl:1997kf,Leupold:1997dg}. 
One still seems to have at least a connection
between chiral restoration --- drop of two- and four-quark condensates --- and
in-medium changes, no matter whether it is a mass shift or a broadening or a more 
complicated in-medium modification \cite{Post:2003hu,Steinmueller:2006id}. 
However, also point 2 and as its consequence point 4 are questionable:
Whether the
four-quark condensates factorize at least in vacuum is discussed since the invention
of QCD sum rules, see, e.g., \cite{Shifman:1978bx,Shifman:1978by,Narison:1983kn,%
Launer:1983ib,Bertlmann:1987ty,Dominguez:1987nw,Gimenez:1990vg,Leinweber:1995fn,%
Bordes:2005wv} and for in-medium
situations also \cite{Eletsky:1992xd,Hatsuda:1992bv,Birse:1996qp,Zschocke:2002mp,%
Leupold:2005eq,Thomas:2005dc}. 
Raising doubts on point 2 immediately
questions point 4 and in that way the seemingly clear connection between chiral
restoration and in-medium changes gets lost.

Indeed, a closer look on the sum rules for light vector mesons reveals that
most of the condensates, whose in-medium change is translated into an in-medium 
modification for the respective hadron, are actually chirally symmetric (see below). 
Physically, it is of course possible and by far not unreasonable that the same 
microscopic mechanism which causes
the restoration of chiral symmetry is also responsible for changes of chirally
symmetric condensates. For example, in the scenario \cite{Park:2005nv} 
about half of
the (chirally symmetric) gluon condensate vanishes together with the two-quark 
condensate.
On the other hand, these considerations show that the connection between the mass
of a light vector meson and chiral symmetry breaking is not as direct as one would
naively expect. 

We take these considerations as a motivation to study in the present work a 
``pure chiral restoration'' scenario, i.e.\ to ask the question: How large would
the mass or the width of the $\rho$ meson be in a world where 
only chiral symmetry breaking objects/condensates are dropped.
We stress that such a scenario
may not reflect all the physics which is contained in QCD. There might be intricate
interrelations between chirally symmetric and symmetry breaking objects.
In that sense the pure chiral restoration scenario shows the minimal impact
that the restoration of chiral symmetry has on the properties of the $\rho$ meson.

 
{\bf 2. Chiral transformations and QCD condensates:}
For vanishing quark masses, QCD with $N_f$ flavors is invariant
with respect to the global chiral $SU_R(N_f) \times SU_L(N_f)$ transformations.
Focusing for the time being on the $N_f = 2$ light (massless) quark sector, the
corresponding left-handed transformations read for the left-handed quark field 
$\psi_L = \frac12 \, (1 - \gamma_5) \, \psi$ and the right-handed quark field
$\psi_R = \frac12 \, (1 + \gamma_5) \, \psi$
\begin{equation}
\psi_L \to e^{i \vec \theta_L \cdot \vec \tau} \psi_L , 
\quad
\psi_R  \to  \psi_R,
\label{eq:chiraltrafos_L}
\end{equation}
while the right-handed transformations are
\begin{equation}
\psi_R \to  e^{i \vec \theta_R \cdot \vec \tau} \psi_R ,
\quad
\psi_L \to  \psi_L,
\label{eq:chiraltrafos_R}
\end{equation} 
where $\vec \tau$ are the iso-spin Pauli matrices and 
$\psi = \genfrac{(}{)}{0pt}{}{u}{d}$
denotes
the quark iso-doublet. Equations (\ref{eq:chiraltrafos_L}, \ref{eq:chiraltrafos_R})
represent isospin transformations acting separately on the right-handed and 
left-handed parts of the quark
field operator $\psi = \psi_L + \psi_R$, i.e.\
the three-component vectors $\vec \theta_{R,L}$ contain
arbitrary real numbers. Gluons and heavier quarks 
remain unchanged with respect to the transformations (\ref{eq:chiraltrafos_L}, \ref{eq:chiraltrafos_R}).

A quark current which has the quantum numbers of the $\rho$ meson is given by the
vector--iso-vector current
\begin{equation}
\label{eq:quarkcurr}
\vec j^\mu = \frac12 \, \bar \psi \gamma^\mu \vec \tau \psi .
\end{equation}
If a chiral transformation according to (\ref{eq:chiraltrafos_L}, \ref{eq:chiraltrafos_R})
is applied to $\vec j^\mu$, it
becomes mixed with the axial-vector--iso-vector current
\begin{equation}
\label{eq:quarkcurr5}
\vec j_5^\mu = \frac12 \, \bar \psi \gamma^\mu \gamma_5 \vec \tau \psi  \, 
\end{equation}
which carries the quantum numbers of
the $a_1$ meson.
 Indeed, experiments show that the vector current (\ref{eq:quarkcurr})
couples strongly to the $\rho$ meson, while the axial-vector current 
(\ref{eq:quarkcurr5}) couples to the
$a_1$ meson \cite{Schael:2005am}. Therefore, $\rho$ and $a_1$ are called chiral partners. 

The central object of QCD sum rules \cite{Shifman:1978bx,Shifman:1978by,Narison}
is the retarded current-current correlator 
which reads for the $\rho^0$ meson
\begin{equation}
  \label{eq:curcur}
  \Pi^{\mu\nu}(q) 
  = i \int\!\! d^4\!x \, e^{iqx} \, \Theta(x_0) \, 
  \langle \left[ j_3^\mu(x) \, , j_3^\nu(0) \right] \rangle \,,
\end{equation}
where for vacuum ($\langle \ldots \rangle$  means accordingly the
vacuum expectation value) the retarded and time-ordered propagator coincide for positive energies, whereas for in-medium situations
(e.g.\ nuclear matter, $\langle \ldots \rangle$ refers then to the Gibbs average), 
the retarded correlator has to be taken (cf.~\cite{Hatsuda:1992bv}).
The imaginary part of the current-current correlator contains the 
spectral distribution, i.e.\ the information which hadronic one-body and many-body states
couple to the considered current. For large space-like momenta,
$Q^2 \equiv -q^2 \gg \Lambda_{QCD}^2$,
the correlator can be reliably calculated from the elementary QCD quark and gluon
degrees of freedom due to asymptotic freedom. 
Results from QCD perturbation theory can be systematically
improved by the introduction of quark and gluon condensates using the operator-product
expansion (OPE) \cite{Wilson:1969zs}. 
The QCD sum rule method  
connects the mentioned two representations of the correlator
by a dispersion relation which reads after a Borel transformation
\begin{equation} \label{sum_rule}
\frac1\pi \int_0^\infty ds \, s^{-1} \, {\rm Im} \, \Pi(s) \, e^{-s/M^2}
= \tilde \Pi (M^2),
\end{equation}
where 
the Borel mass $M$ has emerged from the OPE momentum scale $Q$
(for further details we refer the interested reader 
to \cite{Leupold:1997dg}).
We consider a $\rho$ meson at rest, therefore, the tensor structure of
(\ref{eq:curcur}) reduces to a scalar $\Pi = \frac13 \Pi^\mu _\mu$.
The Borel-transformed OPE reads
\begin{eqnarray} \label{eq:botr}
\tilde \Pi (M^2) &=& c_0 M^2 + \sum_{i=1}^\infty
\frac{c_i}{(i-1)! \,M^{2(i-1)}}
\end{eqnarray}
with coefficients up to mass dimension 6
\begin{eqnarray} \label{coeff1}
c_0 &=& 
\frac{1}{ 8\pi^2}\left(1+\frac{\alpha_s}{\pi} \right) \: ,
\\
c_1 &=& - \frac{3}{ 8\pi^2} (m_u^2 + m_d^2) \: ,
\\ \label{coeff3}
c_2 &=& \frac12 (1 + \frac{\alpha_s}{4 \pi} C_F) 
(m_u \langle \bar u u \rangle + m_d \langle \bar d d \rangle )
+ \frac{1}{ 24} \left\langle \frac{\alpha_s }{ \pi} G^2 \right\rangle + N_2 \: ,
\\ \label{coeff4}
c_3 &=&
-\frac{112}{81} \pi\alpha_s \langle {\cal O}^V_4 \rangle - 4 N_4
\end{eqnarray}
with $C_F = (n_c^2-1)/(2n_c)=4/3$ for $n_c = 3$ colors.
A mass dimension 2 condensate seems to be excluded
in vacuum \cite{Schilcher}.
In (\ref{coeff1} - \ref{coeff4}) we have introduced 
the strong coupling $\alpha_s$, 
the light-quark masses $m_{u,d}$,
the gluon condensate $\left\langle \frac{\alpha_s }{ \pi} G^2 \right\rangle$, 
and the combination of four-quark condensates in compact notation 
\begin{eqnarray}
\langle {\cal O}^V_4 \rangle &= & 
\frac{81 }{ 224} \langle 
(\bar \psi \gamma_\mu \gamma_5 \lambda^a \tau_3 \psi)^2 
\rangle
+ \frac{9 }{ 112} \langle 
\bar \psi \gamma_\mu \lambda^a \psi
\sum\limits_{f = u, d, s} \bar f \gamma^\mu \lambda^a f \rangle
\label{eq:fourqdef}
\end{eqnarray}
with color matrices $\lambda^a$.
(For a classification of four-quark condensates cf.~\cite{Thomas_NPA}. Here, we have extended the notation to the SU(3) flavor sector.)
It is also useful to introduce the averaged two-quark condensate 
$m_q\langle \bar q q \rangle = \frac12 \, \langle m_u\bar u u + m_d\bar d d \rangle$ and $m_q = (m_u+m_d)/2$.
These terms constitute
the contributions which already exist in vacuum (and might change in a medium)
up to higher-order condensates (including, for instance, the poorly known term $c_4$)
which are suppressed by higher powers in the expansion parameter $M^{-2}$. 
Additional non-scalar condensates come into 
play, in particular for in-medium situations.
In (\ref{coeff3}, \ref{coeff4}), only the twist-two non-scalar 
condensates \cite{Hatsuda:1992bv} are displayed,
$
N_i = - \frac23 \, i \, \langle {\cal ST} \bar \psi \gamma_{\mu_1} D_{\mu_2} 
\dots D_{\mu_i} \psi \rangle \, g^{\mu_1 0} \dots g^{\mu_i 0} ,
$
where the operation ${\cal ST}$ is introduced to make the operators symmetric and 
traceless with respect to its Lorentz indices.
Twist-four non-scalar condensates have been found to be numerically 
less important \cite{Hatsuda:1992bv,Hatsuda:1995dy,Leupold:1998bt}.

Using (\ref{eq:chiraltrafos_L}, \ref{eq:chiraltrafos_R}) 
one can show that 
the only objects in the OPE (\ref{eq:botr})
with coefficients (\ref{coeff1} - \ref{coeff4})
which are {\em not} chirally invariant\footnote{Note that $c_1$ breaks the chiral symmetry explicitly. Its contribution is numerically completely negligible.}
are (i) the (numerically small) two-quark condensate 
and
(ii) a part of the (numerically important) four-quark condensate
$\langle {\cal O}^V_4 \rangle$ specified in (\ref{eq:fourqdef}). 
One can split the four-quark condensates (\ref{eq:fourqdef})
into a chirally symmetric part,
\begin{eqnarray}
\langle {\cal O}_4^{\rm sym} \rangle & = &
\frac{81 }{448} \langle 
(\bar \psi \gamma_\mu \gamma_5 \lambda^a \tau_3 \psi)^2 
+ (\bar \psi \gamma_\mu \lambda^a \tau_3 \psi)^2 
\rangle
+ \frac{9 }{ 112} \langle 
\bar \psi \gamma_\mu \lambda^a \psi
\sum\limits_{f = u, d, s} \bar f \gamma^\mu \lambda^a f
\rangle  \,,
\label{eq:4-q-sym}
\end{eqnarray}
and a part which can be transformed into its negative by a proper chiral transformation 
(dubbed ``chirally odd'' object),
\begin{eqnarray}
\langle {\cal O}_4^{\rm br} \rangle & = &
- \frac{81 }{112} \left\langle (\bar \psi_{\rm R} \gamma_\mu \lambda^a \tau_3 \psi_{\rm R})
 \, (\bar \psi_{\rm L} \gamma^\mu \lambda^a \tau_3 \psi_{\rm L}) \right\rangle
  \label{eq:4-q-br}
\end{eqnarray}
with
$\langle {\cal O}^V_4 \rangle = \langle {\cal O}_4^{\rm sym} \rangle + \langle {\cal O}_4^{\rm br} \rangle$.
The last term in \eqref{eq:4-q-sym} is an iso-singlet.
In an isospin invariant system, 
the other terms in (\ref{eq:4-q-sym}) may be written as
$
\left\langle 
(\bar \psi \gamma_\mu \gamma_5 \vec \tau \lambda^a \psi)^2 
+ (\bar \psi \gamma_\mu \vec \tau \lambda^a \psi)^2 
\right\rangle
= 2 \,   \left\langle 
(\bar \psi_R \gamma_\mu \vec \tau \lambda^a \psi_R)^2 
+ (\bar \psi_L \gamma_\mu \vec \tau \lambda^a \psi_L)^2 
\right\rangle
$.
The latter two terms are separately invariant with respect to left-handed and right-handed isospin transformations,
i.e.\ they are chirally invariant. 

{\bf 3. Pure chiral restoration scenario for the rho meson:}
It appears to be very natural that a four-quark condensate which breaks chiral symmetry
can be related to the square of the two-quark condensate which also breaks
chiral symmetry.
Indeed, it has been shown in \cite{Bordes:2005wv} that the factorization of the four-quark condensate $\langle {\cal O}^{\rm br}_{4} \rangle$ given in \eqref{eq:4-q-br} is completely compatible with the ALEPH data on the vector and axial-vector spectral distributions \cite{Schael:2005am}.
On the other hand, it is not so obvious that a chirally symmetric
four-quark condensate like $\langle {\cal O}^{\rm sym}_{4} \rangle$ in (\ref{eq:4-q-sym}) is related
directly to the two-quark condensate.

Along this line of arguments we are going to answer the following question.
What happens to the $\rho$ meson
if one keeps all expectation values of chirally invariant operators at their vacuum
values and puts all chiral symmetry breaking objects to zero? This question defines
what is meant by the ``pure chiral restoration'' scenario. 

\begin{table}
  \centering
  \begin{tabular}{ccc}
    \toprule
    QCD condensate & \hspace{10pt} transformation \hspace{10pt} & vacuum value \\ 
    \midrule
    $\langle \bar q q\rangle$ & chirally odd & $- (240 \, \mbox{MeV})^3$ \\
    $\left\langle \frac{\alpha_s }{ \pi} G^2 \right\rangle$ &  
        invariant & $(330 \, \mbox{MeV})^4$ \\
    $ \langle {\cal O}_4^{\rm sym} \rangle $ & invariant 
        & $ (267 \, \mbox{MeV})^6$ \\
    $ \langle {\cal O}_4^{\rm br} \rangle$ & chirally odd 
       & $\frac97 \, \langle \bar q q \rangle^2$ \\
    \bottomrule
  \end{tabular}
  \caption{Employed QCD condensates, their behavior with respect to chiral transformations 
    and their respective size.}
  \label{tab:tab}
\end{table}

Let us first describe briefly how we fix the numerical values for the QCD
condensates, collected in Tab.~\ref{tab:tab}.
We note that the non-scalar condensates,
which appear in (\ref{coeff1} - \ref{coeff4}), 
are chirally symmetric and vanish in the
vacuum, i.e.\  we can disregard them also for the scenario of pure chiral restoration.
Next we turn to the vacuum condensates.
The gluon condensate is determined from the QCD sum rules for the 
charmonium \cite{Shifman:1978bx,Shifman:1978by}. 
The running coupling has to be evaluated at the
scale $M$. Following \cite{Leupold:2003zb} we use $\alpha_s = 0.38$.
The two-quark condensate is fixed by the Gell-Mann--Oakes--Renner 
relation \cite{GellMann:1968rz}
$m_q \langle \bar q q \rangle  = - \frac12 \, F_\pi^2 \, M_\pi^2  $
with the pion-decay constant $F_\pi \approx 92 \, $MeV, $m_q = (m_u+m_d)/2$ and the pion mass 
$M_\pi \approx 140 \, $MeV \cite{Amsler:2008zzb}. 
Using in addition $m_q = 6 \,$MeV \cite{Narison:1999mv} one gets
$\langle \bar q q \rangle = -(240 \, \mbox{MeV})^3$. 
For vacuum, the condensate $\langle {\cal O}_4^{\rm br} \rangle$ 
has been extracted from the experimental difference
between vector and axial-vector spectral information \cite{Schael:2005am}.
We use the result of \cite{Bordes:2005wv}
\begin{eqnarray}
  \langle {\cal O}_4^{\rm br} \rangle_{\rm vac} & \approx & 
  \frac{9}{7} \, \langle \bar q q \rangle^2_{\rm vac}
  \label{eq:fact-br}  
\end{eqnarray}
together with the vacuum $\rho$-meson properties \cite{Amsler:2008zzb} to 
fix the vacuum value for $\langle {\cal O}_4^{\rm sym} \rangle$ defined 
in (\ref{eq:4-q-sym}). For the pure chiral restoration scenario one drops then
the chiral symmetry breaking terms, in particular (\ref{eq:4-q-br}).
Thus also here one 
only needs the {\em vacuum} values for the chirally invariant terms, 
in particular for (\ref{eq:4-q-sym}).
We recall that \eqref{eq:fact-br} indicates that the chiral symmetry breaking four-quark condensate factorizes \cite{Bordes:2005wv}.

To describe the properties of the $\rho$ meson we rearrange (\ref{sum_rule})
by splitting the integral $\int_0^\infty = \int_0^{s_+} + \int_{s_+}^\infty$
and putting the so-called continuum part to the
OPE terms thus isolating the interesting hadronic resonance part below the
continuum threshold $s_+$. 
This allows to define the normalized moment \cite{Zschocke:2002mn}
of the hadronic spectral function
\eqref{m_bar} 
\begin{eqnarray} \label{m_bar}
\tilde{m}^2 (M,s_+) &\equiv&
\frac{\int_0^{s_+} ds \; {\rm Im} \Pi (s) \; 
e^{-s/M^2}}{\int_0^{s_+} ds \; 
{\rm Im} \Pi (s) \; s^{-1} e^{-s/M^2}}\\ \label{OPE_side}
&=& \frac{c_0 M^2 [1 -(1+\frac{s_+}{M^2}) e^{-s_+ / M^2} ] 
- \frac{c_2}{M^2} - \frac{c_3}{M^4}
- \frac{c_4}{2 M^6}
}
{c_0 [1 - e^{-s_+/M^2}] + \frac{c_1}{M^2} + \frac{c_2}{M^4} + \frac{c_3}{2 M^6}
+ \frac{c_4}{6M^8}
},
\end{eqnarray}
where the semi-local duality hypothesis 
$\frac{1}{\pi} \int_{s_+}^\infty ds s^{-1} {\rm Im} \Pi(s) e^{-s/M^2}
= \int_{s_+}^\infty ds \,c_0 \,e^{-s/M^2}$
is exploited in \eqref{OPE_side}. 
The second line emerges essentially from the OPE and
''measures'' how $\tilde m$ is determined by condensates. 
 
\begin{figure}[t]
\vskip -1cm
\includegraphics[width=0.49\textwidth]{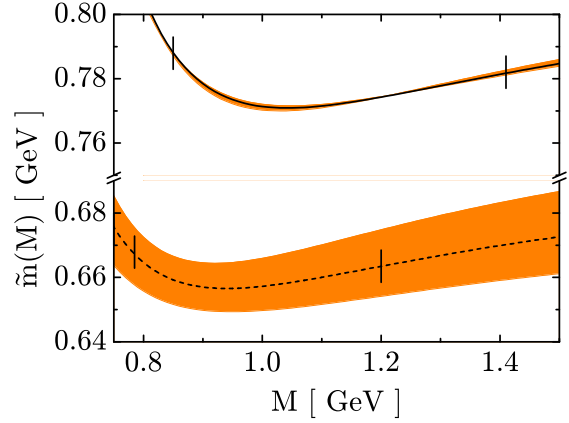}
\caption{The mass parameter $\tilde m (M, s_+)$ for the optimized continuum threshold $s_+$ as a function of the
Borel mass $M$ for vacuum values of condensates from Tab.\ \ref{tab:tab} (upper curves) and the pure chiral restoration scenario with
$\langle {\cal O}^{\rm br}_4 \rangle \rightarrow 0$
and
$\langle \bar q q \rangle \rightarrow 0$ (lower curves).
The Borel windows are marked by vertical bars.
The curves are for $c_4 = 0$, while the bands cover the
range $c_4 = \pm \langle \frac{\alpha_s}{\pi} G^2 \rangle^2$. 
}
\label{fig:Borel}
\end{figure}

The meaning of the spectral moment (\ref{m_bar}) becomes obvious for the pole ansatz \cite{Shifman:1978by} of the
hadronic spectral function below $s_+$, ${\rm Im} \Pi (s) = F_0 \delta(s - m_0^2)$,
where $\tilde{m} = m_0$ follows and $F_0$ is determined by inserting $m_0$ in \eqref{sum_rule}.
For the sake of clarity let us consider first the vacuum case where 
we identify the average of $\tilde{m}$ with the vacuum mass.
The averaged mass parameter is determined by
$\overline{ m} (s_+) = (M_{\rm max} - M_{\rm min})^{-1}
\int_{M_{\rm min}}^{M_{\rm max}} \tilde{m}(M,s_+) \, dM$ 
within the Borel window.
According to \cite{Zschocke} the Borel minimum is determined by the requirement that the mass dimension 6 contribution to the OPE is smaller than 10\%.
For the Borel maximum we demand that the continuum contribution to the spectral integral is smaller than 50\%.
$s_+$ follows from the requirement of maximum flatness of $\tilde{m} (M, s_+) $
as a function of $M$ within the Borel window.
Employing this system of equations,
$\langle \mathcal{O}_4^{\rm sym} \rangle = ( 267 \text{ MeV})^6$ 
with $\langle \mathcal{O}_4^{\rm br} \rangle$ 
from (\ref{eq:fact-br}) (for the other condensates see Tab.~1) is required to get 
$\overline{m} = 775.5$ MeV
for $s_+ = 1.37$ GeV${}^2$, see upper curves in Fig.\ \ref{fig:Borel}.

Let us consider now the pure chiral restoration scenario. 
$\langle \mathcal{O}_4^{\rm br} \rangle \to 0$ and $\langle \bar q q \rangle \to 0$
but keeping the other condensate values
causes a drop of $\overline{m}$
to 659.8 MeV and the continuum threshold becomes $s_+ = 1.03$ GeV${}^2$, see lower curves in Fig.\ \ref{fig:Borel}.
We emphasize the large impact of dropping $\langle {\cal O}^{\rm br}_4 \rangle$ on the averaged spectral moment $\overline m$.

To get an estimate of the possible importance of the poorly known term
$c_4$ we use as ''natural scale'' $\langle \frac{\alpha_s}{\pi} G^2 \rangle^2$.
The Borel curves for $\vert c_4 \vert = \langle \frac{\alpha_s}{\pi} G^2 \rangle^2$
border the bands in Fig.~\ref{fig:Borel}. This estimate is quite rough
as $c_4$ may contain also chirally odd condensates, whose drop is
not accounted for in the pure chiral restoration scenario.  

Summarizing the outcome of this numerical study, the pure chiral restoration
scenario is characterized by a drop of the model-independent spectral moment $\overline{m}$
by about 120 MeV.


{\bf 4. Mass shift vs.\ broadening:}
While for a narrow resonance in vacuum the often employed pole + continuum ansatz 
is reasonable, the spectral distribution may get a more complex
structure in a medium
\cite{Zschocke:2002mn,Steinmueller:2006id,Kwon:2008vq,Friman:1997tc,Rapp:1997fs,Peters:1997va,Post:2003hu,vanHees:2006ng,Klingl:1997kf,Lutz:2001mi}.
In particular, one cannot decide, within the employed framework of QCD sum rules,
whether a drop of $\tilde m$, and consequently $\overline{m}$,
means a mass shift or a broadening or both. 
To make it explicit, we use a Breit-Wigner ansatz
for the spectral function
\begin{equation}
\label{BW}
{\rm Im} \Pi (s\leq s_+) = \frac{F_0}{\pi}
\frac{\sqrt{s} \Gamma(s)}{(s - m_0^2)^2 +s \Gamma^2(s)} \: ,
\end{equation}
where the vacuum parametrization of the width is given by \cite{Leupold:1997dg}
\begin{equation}
\label{Gamma}
\Gamma(s) = \Theta(s-4m_\pi^2) \Gamma_0 
\left(1-\frac{4m_\pi^2}{s}\right)^{\frac 32}
\left(1-\frac{4m_\pi^2}{m_0^2}\right)^{-\frac 32} \: ,
\end{equation}
with $m_\pi = 135$ MeV being the pion mass.
As a consequence of the two-parameter ansatz for the spectral function, the moment $\tilde{m}^2(M)$ determines a relation $m_0(M) = m_0(M, \Gamma_0)$; $F_0(M)$ is again determined by \eqref{sum_rule}, and hence $\overline{m_0} = \overline{m_0}(\Gamma_0)$.

Adjusting $\langle {\cal O}^{\rm sym}_4 \rangle$ to reproduce the experimental mass $\overline{m_0} = 775.5$ MeV and width $\Gamma_0 = 149.4$ MeV we obtain now $\langle {\cal O}^{\rm sym}_4 \rangle = (242 \text{ MeV})^6$.

In general, the peak position $m_{\rm peak}$ and the full width at half maximum $\Gamma_{\rm FWHM}$ of the spectral function do not coincide with the corresponding parameters $m_0$ (or $\overline{m_0}$) and $\Gamma_0$ of the ansatz \eqref{BW}.
For $\Gamma(s) = \Gamma_0 = \text{const.}$, e.g.~, $m_0$ is determined by
$m_0^2 = \sqrt{4 m_{\rm peak}^4 + \Gamma_0^2 m_{\rm peak}^2} - m_{\rm peak}^2$.
While for small $\Gamma_0$ the peak position $m_{\rm peak}$ and $m_0$ differ only by a few MeV,
they differ significantly for larger values of $\Gamma_0$.
Especially, keeping the parameter $\overline{m_0}$ constant in the chiral symmetry restoration scenario causes an implicit shift of the peak position $m_{\rm peak}$ due to the broadening caused by the symmetry restoration.

\begin{figure}[t]
    \vskip -.3cm
    \includegraphics[width=0.49\textwidth]{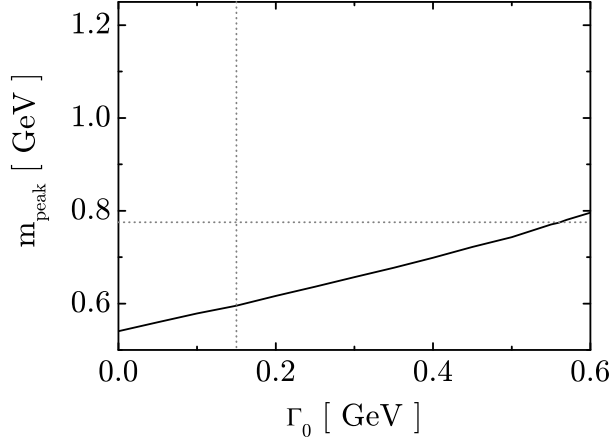}
    \caption{Peak position $m_{\rm peak}$ as a function of the width parameter $\Gamma_0$ for $\langle {\cal O}^{\rm sym}_4 \rangle = (242 \text{ MeV})^6$.
    Dotted lines mark the experimental values $m_{\rm peak} = 775.5$ MeV and $\Gamma_{\rm FWHM} = 149.4$ MeV.}
    \label{fig:shape}
\end{figure}

Therefore, instead of requiring one of the possible options $\overline{m_0}$ or $\Gamma_0$ to be constant, we now demand that the associated shape characteristics of the spectral function, i.e. $m_{\rm peak}$ or $\Gamma_{\rm FWHM}$, are fixed within the restoration procedure.
In doing so, it is convenient to set $\Gamma(s) = \Gamma_0 = \text{const}$ for the chiral symmetry restoration scenario.
The result is depicted in Fig.\ \ref{fig:shape}.
For the pure chiral restoration scenario, the curve $m_{\rm peak}(\Gamma_0)$ is significantly 
shifted away from the vacuum physical point $(m_0, \Gamma_0)$ = (775.5 MeV, 149.4 MeV).
If one assumed that chiral restoration in the present spirit does not cause an additional broadening, one would recover the previously often anticipated ''mass drop``.

Fig.~\ref{fig:shape} evidences, however, that an opposite interpretation is conceivable as well, namely pure broadening with keeping the vacuum value of $m_{\rm peak}$.
The NA60 \cite{Arnaldi:2006jq,Adamova:2006nu} and CLAS \cite{CLAS} 
data seem indeed to favor such a broadening
effect.
In fact, assuming that $m_{\rm peak}$ does not change by chiral restoration, the width is increased to 600 MeV.
In this respect, broadening of a spectral function signals equally well chiral
restoration as dropping mass would do.

\begin{figure}[t]
    \vskip -1cm
    \includegraphics[width=0.49\textwidth]{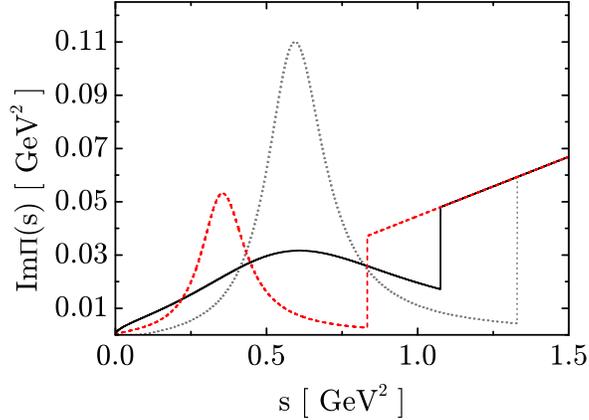}
    \caption{The spectral density $\text{Im} \Pi(s)$ in the vacuum case (dotted curve) and in the pure chiral restoration scenario for $\Gamma_{\rm FWHM} = \text{const}$ (dashed curve) or $m_{\rm peak} = \text{const}$ (solid curve).}
    \label{fig:shapeb}
\end{figure}

Fig.\ \ref{fig:shapeb} exhibits the spectral function ${\rm Im} \Pi(s)$ as a function of $s$ for the two extreme options above.
The solid curve depicts the enormous broadening when keeping the peak at the vacuum position.
The dashed curve is for the dropping mass option for keeping the full width at half maximum at its vacuum value.
From the perspective of the employed QCD sum rules, both options are equivalent, as any other point on the curve $m_{\rm peak}(\Gamma_0)$ in Fig.\ \ref{fig:shape}.

The overall outcome seems to be that the Borel transformed QCD sum rule requires more strength of the spectral function at lower energies.
This may be realized by a shift or a broadening or both.

It should be emphasized, in this context, that (\ref{BW}) is a pure ad hoc ansatz. 
For instance, multi-peak structures may emerge from particle-hole and resonance-hole excitations in the nuclear medium \cite{Randrup,Rapp:1997fs,Peters:1997va,Post:2003hu,Lutz:2001mi}.
It is desirable, therefore, to determine the spectral distribution
by experimental data of $\rho$ decay into dileptons in a nuclear medium
to decide whether it deviates from the vacuum
and try to relate it to the change of condensates and Landau term. 
Such strategy for a particular model instead of data is also envisaged in \cite{Kwon:2008vq} in a truncated hierarchy of spectral moments.
Clearly a better microscopic understanding is mandatory for an appropriate modelling of the spectral shape.
This applies also for the continuum threshold region.

{\bf 5. Notes on omega and axial-vector mesons:}
The current 
$\bar u \gamma^\mu u + \bar d \gamma^\mu d$ has the quantum numbers of the $\omega$ meson
and is a chiral singlet with respect to $SU_R(2) \times SU_L(2)$ chiral transformations.
Consequently, the $\omega$ meson does not have a chiral partner {\em in a world with
two light flavors}. Concerning {\em three} light flavors the current given above
is a superposition of a member of the flavor octet and the singlet. The octet members
do have chiral partners and one may assign a proper linear combination of the two 
$f_1$ \cite{Amsler:2008zzb} mesons as the chiral partner of the respective linear combination of $\omega$ and
$\phi$. 

All considerations made in the following 
concern {\em two} light flavors. The OPE side
of the $\omega$ meson including terms up to dimension 6 contains only chirally
symmetric terms \cite{Shifman:1978bx,Shifman:1978by,Hatsuda:1992bv} -- except for the two-quark condensate term 
$\propto m_q \langle \bar q q \rangle$. This term, however, breaks chiral symmetry explicitly
by the quark mass and dynamically by the quark condensate. Considerations
about symmetry transformations, on the other hand, concern the case where chiral symmetry
is exact, i.e.\ without explicit breaking. Therefore the appearance of the term
$\propto m_q \langle \bar q q \rangle$ is not in contradiction to the 
statement that the $\omega$ is a chiral singlet. Without explicit calculations it is clear
that in the pure chiral
restoration scenario the $\omega$ meson does not change much of its mass since
only the numerically very small term $\propto m_q \langle \bar q q \rangle$ is dropped
while the chirally symmetric four-quark condensates do not change.

The current (\ref{eq:quarkcurr5}) with the quantum numbers of the $a_1$ meson
yields the same chirally symmetric OPE parts as the $\rho$ meson. The chirally odd parts
are of course different, they are the negative of the ones which appear in the OPE for
the $\rho$ meson \cite{Shifman:1978bx,Shifman:1978by,Hatsuda:1992bv}.
In addition, the hadronic side of the sum rule contains not
only the $a_1$, but also the pion. The latter contribution is $\propto F_\pi^2$ where
$F_\pi$ denotes the pion-decay constant. Both effects, different chirally odd condensates
and the appearance of the pion, lead to the fact that the sum rule method yields
a mass for the $a_1$ which is significantly different from the 
$\rho$ meson mass \cite{Shifman:1978bx,Shifman:1978by} --
as it should be. In the pure chiral restoration scenario the chirally odd condensates
are put to zero. In addition, the pion-decay constant which is an order parameter
of chiral symmetry breaking \cite{Meissner:2001gz} also vanishes. 
Then the sum rules for $\rho$ and $a_1$ are the same. As expected the chiral partners 
become degenerate in the pure chiral restoration scenario. 

{\bf 6. Summary:}
Two extreme and antagonistic statements concerning hadron masses and hadronic in-medium
modifications could be raised:
(a) Basically all hadron masses are caused by chiral symmetry breaking.
  Consequently in a dense and/or hot strongly interacting medium the masses of hadrons 
  vanish at the point of chiral restoration -- apart from some small remainder which is
  due to the explicit breaking of chiral symmetry by the finite quark masses.
(b) The observed in-medium changes can be explained by standard hadronic 
  many-body approaches and have
  no direct relation to 
  chiral restoration.
Our findings do not support either of these extreme statements. If one drops the
chiral symmetry breaking condensates in the sum rule for the $\rho$ meson one does
see a significant change of the mass moment.
This is neither 100 \% 
(as statement (a) would suggest) nor 0 \% (statement (b)).
In the scenario of pure
chiral restoration we have kept the chirally invariant condensates at their vacuum
values.
Though, an adequate restoration mechanism might also change these condensates, this allows a discussion of chiral-symmetry restoration which is not interfered by additional in-medium effects.
This is clearly unrealistic for a true in-medium situation.
In particular,
the contributions coming from the non-scalar twist-two operators
are found to be sizable, e.g., for cold nuclear 
matter \cite{Hatsuda:1991ez}. Nonetheless, our scenario
indicates that the connection between the vacuum masses and chiral symmetry breaking
or between dropping masses and chiral restoration
is not direct. In principle, one could imagine conspiracies between chiral symmetry 
breaking and 
non-breaking condensates such that one of the extreme statements raised above becomes
true. Such a conspiracy would be driven by the underlying microscopic mechanisms which
cause spontaneous chiral symmetry breaking and/or its restoration. Clearly
we need a deeper understanding of these microscopic mechanisms.
Finally, we emphasize that, within the framework of QCD sum rules,
a ``dropping mass'' and a broadening of the spectral function
are both equally well conceivable in their relation to chiral restoration.

\acknowledgments
The work of S.L. has been supported by GSI Darmstadt.
T.H., R.T., and B.K. acknowledge the support by GSI-FE and BMBF 06DR9059.

\end{document}